\documentclass{article}
\title{A comparison of Zeroes and Ones of a Boolean Polynomial}
\author{M.~N.~Vyalyi\thanks{Supported by RFBR grant 99-01-00122}\\
vyalyi@mccme.ru}
\date{November 20, 2001}

\usepackage{amsfonts,amssymb,amsmath,amsthm}
\usepackage{fullpage}
\newtheorem*{Remark}{Remark}
\newtheorem*{Theorem}{Theorem}
\newtheorem{theorem}{Theorem}
\newtheorem{lemma}{Lemma}

\let\epsilon\varepsilon

\let\al\alpha

\let\ph\varphi

\def\CC{\mathbb C}

\def\ZZ{\mathbb Z}

\def\FF{{\mathbb F}_2}
\def\cb{{\mathbb B}}

\def\bydef{\mathrel{\stackrel{\scriptscriptstyle\text{def}}{=}}}

\renewcommand*\P{\ensuremath{\mathrm {P}}}
\newcommand*\PP{\ensuremath{\mathrm {PP}}}

\newcommand*\BQP{\ensuremath{\mathrm {BQP}}}
\newcommand*{\poly}{\mathop{\mathrm{poly}}}
\newcommand*\GapP{\ensuremath{\mathrm {GapP}}}

\def\ket#1{|#1\rangle }
\def\bra#1{\langle #1|}

\def\sx{\sigma_x}
\def\sy{\sigma_y}
\def\sz{\sigma_z}

\def\SU{\mathop{\mathbf{SU}}}

\def\monom{\mathop{\mathrm{M}}}
\def\Re{\mathop{\mathrm{Re}}}

\def\Card{\mathop{\mathrm{Card}}}

\def\Fy{F_{\mathrm{yes}}}
\def\Fn{F_{\mathrm{no}}}

\begin{document}
\maketitle

\begin{abstract}
In this paper we consider the computational complexity of the
following problem. Let $f$ be a Boolean polynomial.
What value of~$f$, 0 or 1, is taken more frequently?
The problem is solved in polynomial time for polynomials of degrees 1,
2.  The next case of degree~3 appears to be \PP-complete under
polynomial reductions in the class of promise problems.  The proof is
based on techniques of quantum computation.
\end{abstract}

The class \PP{} was defined by J.\,Gill~\cite{Gill} as probabilistic
polynomial time with unbounded error. The following problem
represents all computational power of \PP: \emph{A Boolean function
$f$ is given by a Boolean circuit computing this function.
What value of~$f$, 0 or 1, is taken more frequently?} In fact, this
\emph{value comparison problem} is \PP-complete under polynomial
reductions, if \emph{promise problems} are considered instead of
decision problems.  A promise problem  is a decision problem in which
some inputs are excluded. So, a promise problem $F$ is described by a
pair of disjoint sets $(\Fy,\Fn)$ of strings corresponding to ``yes''
and ``no'' instances.

In this paper we restrict the value comparison
problem to the case of Boolean polynomials
of fixed degree. It is known that the problem of
counting of zeroes for polynomials of degree~3 is
\#\P-complete~\cite{EK}.  An easy corollary of this result is
\PP-completeness of the value comparison problem
for polynomials of degree~4 (see
Theorem~\ref{4sgnDf=>PPcomplete} below).

At other hand, the comparison problem for polynomials of degree 1
is trivial. In the case of degree~2 the problem is solved in polynomial
time (by reduction to the canonical form, see~\cite{McWSl, LN}).

We address to the remaining case --- polynomials of degree~3.
It will be shown that it is \PP-complete. Surprisingly enough, the
proof will use techniques of quantum computation.  We will apply the
theorem of efficient approximation for unitary operators~\cite{Kit97}
and the results of~\cite{KnillLaflamme}.  It was shown
in~\cite{KnillLaflamme} that a problem of determination of sign of
specific quadratically signed weight enumerators is \BQP-complete
(again, we mean the completeness in the class of promise problems).  It
is possible to use the results of~\cite{KnillLaflamme} directly for
the proof of our main theorem. Instead, we prefer to
follow the arguments of~\cite{KnillLaflamme} and present a slightly
more restrictive form of the enumerators.

\section{Preliminaries}

\subsection{An another definition of\boldmath{} \PP}

We will also use the definition of the class
\PP{} given by Fenner, Fortnow and Kurtz~\cite{FFK94}.
They introduced the class
\GapP{} functions consisting of the closure under substraction of the
set of \#\P{} functions. In other words,
for any $\GapP$ function $f\colon\cb^*\to\ZZ$ there are predicates
$Q_1(\cdot,\cdot),Q_2(\cdot,\cdot)\in \P$ and a polynomial
$q(\cdot)$ such that for all~$x$
\begin{equation} f(x)=\Card\{y : Q_1(x,y) \& |y|=q(|x|)\}-
\Card\{y : Q_2(x,y) \& |y|=q(|x|)\}.
\end{equation}
The class \PP{} can be defined in these terms as follows:

\emph{The class \PP{} consists of those promise problems $F$ such that
for some \GapP{} function $f$  \emph{(an indicator function)} and all
$x$
\begin{equation} x\in \Fy \Longrightarrow f(x)>0, \qquad x\in \Fn
\Longrightarrow f(x)<0.
\end{equation}
}

\subsection{Notations and simple facts about Boolean polynomials}

\emph{A Boolean polynomial} is a polynomial over the field~$\FF$
consisting of two elements.

Let $\#_0 f$ be the number of zeroes for a Boolean polynomial~$f$,
$\#_1f$ the number of ones. It is clear that
$\#_0f+\#_1f=2^n$. The difference of these numbers will be denoted
by $\Delta f$:
\begin{equation}\label{defDelta}
\Delta f=\#_0f-\#_1f=\sum_{x\in\FF^n}^{}(-1)^{f(x)}.
\end{equation}
In the case $\Delta f=0 $ the polynomial will be called
\emph{balanced}.

So, the value comparison problem can be reformulated as the problem of
determination of the sign of~$\Delta f$.

We shall assume throughout the paper that polynomials are represented
in the form of monomial sum.

Now we introduce some simple properties of~$\Delta f$.

Let $f|_L$ be
the restriction of the polynomial~$f\in\FF[x_1,\dots,x_n]$
to the subspace~$L$ of~$\FF^n$ (hereinafter we will consider  affine
subspaces of~$\FF^n$).

\begin{lemma}\label{Dneg}
$\Delta(1+f)=-\Delta f$.
\end{lemma}

\begin{lemma}\label{Dsum}
If $\ph\colon\FF^n\to\FF$ is a non-zero linear functional, then
$\Delta f=\Delta f|_{\ph(x)=0}+\Delta f|_{\ph(x)=1}$.
\end{lemma}

By direct use of the second equation in~\eqref{defDelta}, we get the
following lemma.

\begin{lemma}\label{Dprod}
If $f(x,y)=g(x)+h(y)$, then $\Delta f = \Delta g \cdot\Delta h$.
\end{lemma}

\begin{lemma}\label{lin}
If $\ell\colon\FF^n\to\FF$ is a
non-zero linear functional on~$\FF^n$,
then $\Delta\ell=0$.
\end{lemma}

\begin{proof}
Changing a basis, we transform
$\ell$ into the form $\ell(x)=x_1$. For this functional the statement
is obvious.
\end{proof}

\begin{lemma}\label{Dsubspace}
Suppose a subspace~$L$ is given by a system of equations
$\ell_i(x)=0$, $1\leq i\leq d$.
Define the polynomial~$g(x,v)$ of
$n+d$ variables by the formula $g=f+\sum_{j=1}^{d}v_j\ell_j(x)$. Then
\begin{equation}\label{eqDsub}
\Delta g=2^d\Delta f|_L.
\end{equation}
\end{lemma}

\begin{proof}
Let us consider  polynomials $g_x(v)=g(x,v)$.
Lemmata~\ref{Dprod},~\ref{lin} imply that for any $x\notin L$
the polynomial $g_x$ is balanced. So, it
contributes 0 to the $\Delta g$.
For any $x\in L$ the polynomial
$g_x$ is not depend on values of~$v_j$. Hence, it
contributes $2^d(-1)^{f(x)}$ to the $\Delta g$. Summing contributions
over all $x\in\FF^n$, we get~\eqref{eqDsub}.
\end{proof}

In the proof of \PP-completeness a relation between quadratically
signed weight enumerators and $\Delta f$ will be exploited. Namely,
consider a quadratically signed weight enumerator
\begin{equation}\label{12-sum}
S(A,B)=\sum_{Ax=0, x\in\FF^n}   (-1)^{B(x)} 2^{|x|}4^{n-|x|},
\quad \deg B=2,
\end{equation}
where $A$ is a Boolean matrix, $|x|$ is the number of ones in
$\{x_j\}_{j=1}^n$.  Let $a_{jk}$ are matrix elements of~$A$. By
$f_{A,B}$ denote the polynomial of $4n$ variables:
\begin{equation}\label{q-pol}
f_{A,B}(x,y,z,u)=\sum_{j=1}^{â}x_jy_jz_j+B(x)
+\sum_{k=1}^{n}u_k\sum_{j=1}^{n}a_{ij}x_j.
\end{equation}

\begin{lemma}\label{QWEtoDelta}
$S(A,B)=2^{-n}\Delta f$.
\end{lemma}
\begin{proof}
We calculate
$\Delta f$ for each $x$ separately. If
 $Ax\ne0$, then $f$ is reduced to
$\ell(u)+g(y,z)$, $\deg \ell=1$, and $\Delta(\ell(u)+g(y,z))=0$.
If $Ax=0$, then $f$ is reduced to
$f_x=B(x)+\sum_{j: x_j=1}y_jz_j$ and does not depend on~$u$.
In this case we get
\begin{equation}\label{Ax=0}
\Delta f_x=2^n\sum_{y,z}^{}(-1)^{f_x(y,z)}=(-1)^{B(x)}2^n4^{n-|x|}
\prod_{j:x_j=1}\sum_{y_j,z_j\in\FF}(-1)^{y_jz_j}
=2^n(-1)^{B(x)}2^{|x|}4^{n-|x|}.
\end{equation}
Summing~\eqref{Ax=0}, we obtain $\Delta f=2^{n}S(A,B)$.
\end{proof}

\subsection{Some facts about quantum computation}

Basics of quantum computation can be found in~\cite{Kit97,BV97}.
Here we recall two facts that are used in the
following proof.

The main tool will be the theorem on efficient approximation of unitary
operators~\cite{Kit97}.
\begin{Theorem}
Let elements
$X_1,\dots,X_l\in\SU(n)$ generate an everywhere dense set in
$\SU(n)$. There is an algorithm which construct for any
matrix
$U\in\SU(n)$ and any precision threshold~$\delta$ an
$\delta$-approximation~$\tilde U$ of~$U$ in the form of product of
generators and their inverses $X_1,\dots,X_l,X_1^{-1},\dots,X_l^{-1}$.
The algorithm runs in time $\exp(O(n)\poly\log(1/\delta))$.

\upshape
$\delta$-approximation means that
$\|U-\tilde U\|<\delta$ in operator norm.
An implicit factor in the running time bound may depend on
$X_1,\dots,X_l\in\SU(n)$.
Matrix elements of~$U$ may be any efficiently computable complex
numbers.
\end{Theorem}

We will approximate by a set of unitary operators
$\exp(i\ph\sigma(s))$, $\|s\|\leq2$, $\ph$ is real.
In fact, we will need $\ph=\arccos(2/\sqrt5)$ only.
Here the following notations are used
\begin{equation}
\sigma(s)=
\sigma(\alpha_1,\beta_1, \alpha_2,\beta_2, \dots, \alpha_n,\beta_n)
\bydef
\sigma_{\alpha_1,\beta_1}\otimes\sigma_{\alpha_2,\beta_2}\otimes \dots
\otimes\sigma_{\alpha_n,\beta_n},\qquad s\in \FF^{2n},
\end{equation}
$\sigma_{\alpha_1,\beta_1}$ are \emph{Pauli matrices}:
\begin{align*}
&\sigma_{00}=
\begin{pmatrix}1&0\\0&1\end{pmatrix}\quad
&&
\sigma_{01}=
\begin{pmatrix}1&0\\0&-1\end{pmatrix}=\sz;\quad
&&\sigma_{10}=
\begin{pmatrix}0&1\\1&0\end{pmatrix}=\sx;\quad
&&
\sigma_{11}=
\begin{pmatrix}0&-i\\ i&0\end{pmatrix}=\sy.
\end{align*}
A weight
$\|s\|$ is equal to the number of non-zero pairs
$\al_j,\beta_j$ in $s$.

The following lemma is due to~\cite{KL98, KLZ}.

\begin{lemma}\label{dense}
Suppose $\ph$ is incommensurable with $\pi$; then operators
$\exp(i\ph\sigma(s))$ generate an everywhere dense set in
$\SU((\CC^2)^{\otimes 2})$.
\end{lemma}

\medskip

{\textit{Sketch of proof}. (See~\cite{Kit97,KShV} for
details.)  The operators $\exp(i\ph\sz)$ and $\exp(i\ph\sx)$ do not
commute.  Hence, the operators
$\exp(i\ph\sigma(1,0,0,0))\exp(i\ph\sigma(1,0,0,1))$
and
$\exp(i\ph\sigma(0,1,0,0))\exp(i\ph\sigma(0,1,0,1))$
generate an everywhere dense set in $\SU(\CC^2\otimes\ket{0})$.
Similarly,
the operators
$\exp(i\ph\sigma(0,0,1,0))\exp(i\ph\sigma(0,1,1,0))$
and
$\exp(i\ph\sigma(0,0,0,1))\exp(i\ph\sigma(0,1,0,1))$
generate an everywhere dense set in $\SU(\ket{0}\otimes\CC^2)$.
Multiplying these sets, we obtain an everywhere dense set
in $\SU(\CC(\ket{00})\oplus\CC(\ket{01})\oplus\CC(\ket{10}))$.
To complete the proof it remains to
note that the operator $\exp(i\ph(0,0,1,0))$ does not fix the subspace
$\CC(\ket{11})$.
\hfill$\square$\parfillskip0pt\par}

\section{The case of polynomials of degree~4}\label{deg4}

\begin{theorem}\label{4sgnDf=>PPcomplete}
The value comparison problem for polynomials of degree~4 is
\PP-complete.

\upshape
(The problem is considered as a promise problem and the class \PP{}
is assumed consisting of promise problems. A reduction is a polynomial
reduction in the class of promise problems.)
\end{theorem}

\begin{proof}
Let $F\in\PP$ be a promise problem,
$f\in\GapP$ its indicator function, and
 $Q_1(x,y),Q_2(x,y)\in\P$ predicates from the
definition of the class \GapP{} applied to~$f$. On inputs $x$ of
length~$n$ the predicates $Q_j$ are computed by polynomial size
Boolean circuits over the basis $\{\cdot,+\}$. Adding dummy assignments
if necessary, we may assume that both circuit sizes are equal to
$s=\poly(n)$.

By $z^{(j)}_k$, $1\leq k\leq s$, we denote auxiliary variables of the
circuit computing the predicate~$Q_j$.  We also assume that the value
of the circuit is the value of the variable $z^{(j)}_{s}$. Each
assignment in a circuit has the form $z^{(j)}_k:=a*b$ where
$*\in\{+,\cdot\}$ and $a,b$ are either input or auxiliary variables.
The equation $Z^{(j)}_k=z^{(j)}_k+a*b=0$ coressponds to
this assignment. Note that the values of input variables $x,y$
determine the values of all auxiliary variables. So, for each $x$ the
number of solutions of the system of equations $Z^{(j)}_k=0$, $1\leq
k<s$, $z^{(j)}_{s}=1$ equals $\Card\{y :  Q_j(x,y) \& |y|=q(|x|)\}$. At
other hand, by the argument of Lemma~\ref{Dsubspace} this number
equals $2^{-s}\Delta F^{(j)}_x$, where
$$F^{(j)}_x(y,z,v)=\sum_{k=1}^{s}v_kZ^{(j)}_k+v_0(z^{(j)}_{s}+1).$$

Therefore, we get $f(x)=2^{-s}(\Delta F^{(1)}_x-\Delta F^{(2)}_x)$.
Taking into account the relation $(\Delta F^{(1)}_x-\Delta
F^{(2)}_x)=\Delta((1+w)F^{(1)}_x+w(1+F^{(2)}_x))$, we obtain the
reduction $x\mapsto F_x$, where
\begin{equation}\label{PPcomplete}
F_x=(1+w)F^{(1)}_x+w(1+F^{(2)}_x).
\end{equation}
It is clear that this reduction is polynomial.
\end{proof}

\section{The case of polynomials of degree~3}\label{deg3}

\begin{theorem}\label{3sgnDf=>PPcomplete}
The value comparison problem for polynomials of degree~3 is
\PP-complete.
\end{theorem}

\begin{proof} We will use the Theorem~\ref{4sgnDf=>PPcomplete}.
So, we will construct for any polynomial $f$, $\deg f=4$, a
polynomial~$g$, $\deg g=3$, such that the signs of
$\Delta f$ and $\Delta g$ are equal.
The construction should be done in time
polynomial of the input size (polynomial of $n$, where $n$ is the
number of variables of~$f$).

At first, we define a unitary operator~$U(f)\colon(\CC^2)^{\otimes n}
\to(\CC^2)^{\otimes n}$ by the following way.
The operator $S_J=\Lambda^J(-1)$
(controlled phase shift)
 corresponds to the monomial
$x_J=\prod_{j\in J}x_{j}$ of~$f$. Controlled phase shift is defined as
\begin{equation}
\left\{\begin{array}{ll}
\Lambda^J(-1)\ket{x_1\dots x_n}=-\ket{x_1\dots x_n},\qquad
&x_J=1,\\
\Lambda^J(-1)\ket{x_1\dots x_n}=\ket{x_1\dots x_n},\qquad
&\text{otherwise}.
\end{array}\right.
\end{equation}

Let
\begin{equation}\label{U(f)}
U(f)=\prod_{j=1}^{n}H[j] \prod_{J\in\monom(f)}S_J\prod_{j=1}^{n}H[j],
\end{equation}
where $H$ is the Hadamard matrix
$$ H=\frac{1}{\sqrt2} \begin{pmatrix}1&1\\1&-1 \end{pmatrix}$$
and $\monom(f)$ is the set of the monomials of~$f$.

We have
\begin{multline}\label{0U0}
\bra0U(f)\ket0=\\=\frac{1}{2^n}\sum_{x_1,\dots,x_n}
\bra{x_1,\dots,x_n} \prod_{J\in\monom (f)}S_J\ket{x_1,\dots,x_n}=
\frac{1}{2^n}\sum_{x_1,\dots,x_n}(-1)^{\sum_{J\in\monom(f)}x_J}=
\frac{1}{2^n}\sum_{x_1,\dots,x_n}(-1)^{f(x)}=\\=2^{-n}\Delta f.
\end{multline}

Note that
$\det U(f)=\pm1$. W.l.o.g. we assume that
$U(f)\in\SU((\CC^2)^{\otimes n})$. (The case of $\det U(f)=-1$
is essentially the same.)

Now, we approximate
$U(f)$ with precision $\delta=2^{-n-1}$ by an operator
$\tilde U$ in the form of product of the operators
$\exp(i\ph\sigma(s))$, $\|s\|\leq2$, where $\cos\ph=2/\sqrt5$.
To achieve the precision required we approximate each factor
in~\eqref{U(f)} with greater precision~$O(\delta/n^4)$.  Each factor
in~\eqref{U(f)} acts on 4 bits at most.  So, by the theorem on
efficient approximation an operator $\tilde U$ can be constructed in
$\poly(n)$ time.  Assume that
\begin{equation}\label{exp} \tilde U=
\prod_{j=1}^{N}\exp(i\ph\sigma(s_j),\qquad N=\poly(n),\
s_j=(\al_{j1},\beta_{j1},\dots,\alpha_{jN},\beta_{jN}).
\end{equation}

From
$
|\bra0U(f)-\tilde U\ket0|\leq \|U(f)-\tilde U\|<\delta
$ and~\eqref{0U0} we conclude that the sign
$\Re \bra0\tilde U\ket0$ equals the sign of
$\Delta f$.

The next step is to find out
$\bra0\tilde U\ket0$:
\begin{multline}\label{0tU0}
 \bra0\tilde U\ket0=
\bra0\prod_{j=1}^{N}\exp(i\ph\sigma(s_j))\ket0=
\bra0\prod_{j=1}^N(\cos\ph+i\sin\ph\sigma(s_j))\ket0= \\=
\sum_{x_1,\dots,x_n}\prod_{x_j=0}\cos\ph
\bra0\prod_{x_j=1}i\sin\ph\sigma(s_j)\ket0=
\sum_{x_1,\dots,x_N}\prod_{x_j=0}\cos\ph\prod_{x_j=1}i\sin\ph
\bra0\prod_{x_j=1}\sigma(s_j)\ket0.
\end{multline}

An operator
$\sigma(s_j)$ flips a bit $k$ iff $\al_{jk}=1$. Hence,
if  $Ax\ne0$, where  $A_{jk}=\al_{jk}$, then
$(x_1,\dots,x_N)$ contributes a zero to the sum~\eqref{0tU0}.

Let us evaluate a phase factor of
 $\bra0\prod_{x_j=1}\sigma(s_j)\ket0$.
Let $y_{jk}$ be the value of the bit $k$  before application of
$\exp(i\ph\sigma(s_j)$. By direct calculation we get
$y_{jk}=\sum_{t=1}^{j-1}\al_{tk}x_t$.
By $\gamma_j$ denote the number of $\sy$ operators in
$\sigma(s_j)$. Then
$\sigma(s_j)$ multiply a phase by a factor
\begin{equation}\label{phase}
i^{\gamma_j}\prod_{k=1}^{n}(-1)^{\beta_{jk}y_{jk}}.
\end{equation}
We put~\eqref{phase} into~\eqref{0tU0} and obtain
\begin{multline}\label{0tU0'}
 \bra0\tilde U\ket0=
\sum_{Ax=0}(\cos\ph)^{N-|x|}(\sin\ph)^{|x|}
i^{\sum(\gamma_j+1)x_j}
\prod_{j=1}^{N}\prod_{k=1}^{n}(-1)^{x_j\beta_{jk}y_{jk}}
=\\=
\sum_{Ax=0}(\cos\ph)^{N-|x|}(\sin\ph)^{|x|}
i^{\sum(\gamma_j+1)x_j}
(-1)^{\tilde B(x)}=\frac{1}{5^{N/2}}
\sum_{Ax=0}2^{N-|x|}i^{\sum(\gamma_j+1)x_j}(-1)^{\tilde B(x)}.
\end{multline}
Let us introduce the notation $\Gamma_t=\{j:\gamma_j=t\}$.
Using the obvious identity
$$ x_1\oplus x_2\oplus\dots\oplus x_r=
x_1+\dots+x_r-2s_2(x_1,\dots,x_r) \pmod4,\quad (x_i\in\{0,1\}),
\qquad s_2(x_1,\dots,x_r)=\sum_{j\ne k}x_jx_k,
$$
we rewrite the power of $i$ in~\eqref{0tU0'} as
\begin{equation}\label{eq3}
i^{\sum(\gamma_j+1)x_j}=(-1)^{\sum_{j\in\Gamma_1}^{}x_j}
(-1)^{s_2(x_{\Gamma_0})+s_2(x_{\Gamma_2})}
i^{\oplus_{j\in\Gamma_0}x_j-\oplus_{j\in\Gamma_2}x_j},
\end{equation}
where $x_{\Gamma_t}$ is the set of variables $x_j$ whose indexes are
in~$\Gamma_t$.

Thus, $\Re \bra0\tilde U\ket0$ can be expressed in the form~\eqref{12-sum}.
The quadratic weight in this representation is
$B(x)=\tilde B(x)
+s_2(x_{\Gamma_0})+s_2(x_{\Gamma_2})$ while the subspace is given by
equations $Ax=0$, $gx=0$ ($g_j=1$ iff $\gamma_j$ is even):
\begin{equation} \Re \bra0\tilde U\ket0=
\frac{1}{5^{N/2}}\sum_{Ax=0,
gx=0}2^{N-|x|}(-1)^{B(x)}=
\frac{1}{20^{N/2}}\sum_{Ax=0,gx=0}2^{|x|}4^{N-|x|}(-1)^{B(x)}.
\end{equation}
By Lemma~\ref{QWEtoDelta}, up to a positive factor this expression
equals $\Delta g$ for some polynomial $g$ of degree 3. It follows from
the proof of Lemma~\ref{QWEtoDelta} and the construction above that
the polynomial~$g$ can be built in polynomial time from the
representation of~$\tilde U$ in the form~\eqref{exp}.
\end{proof}

\begin{Remark}\upshape
The use of Theorem~\ref{4sgnDf=>PPcomplete} is not necessary.
It would be enough to note that the computation of any
predicate from the class $\P$ on inputs of length $n$ can be done
by a reversible circuit of size $\poly(n)$. The rest of proof
remains the same.
\end{Remark}

\end{document}